\documentclass[aps,prb,preprint,superscriptaddress,showkeys]{revtex4}

\usepackage{textcomp}
\usepackage{amsmath}
\usepackage{amssymb}
\usepackage{graphicx}

\newcommand{\fbeta}{f_\beta \! \left( \beta \right)}
\newcommand{\jmmapprox}{\sim \! \!}
\newcommand{\jmmemph}[1]{\textit{#1}}

\usepackage{color}

\begin{document}

\title{The Combined Influence of Nuclear Quantum Effects and van der Waals Interactions on the Structure of Ambient Water}

\author{Jeffrey M. McMahon}
\email[]{mcmahonj@illinois.edu}
\affiliation{Department of Physics, University of Illinois at Urbana-Champaign, Illinois 61801, USA}

\author{Miguel A. Morales}
\affiliation{Lawrence Livermore National Laboratory, Livermore, California 94550, USA}

\author{Brian Kolb}
\affiliation{Department of Physics, Wake Forest University, Winston-Salem, North Carolina 27109, USA}

\author{Timo Thonhauser}
\affiliation{Department of Physics, Wake Forest University, Winston-Salem, North Carolina 27109, USA}


\date{\today}

\begin{abstract}
Path-integral molecular dynamics simulations based on density functional theory employing exchange--correlation density functionals capable of treating nonlocal van der Waals (vdW) interactions self-consistently provide a remarkably accurate description of ambient water. Moreover, they suggest that water's structure may be impacted by a \jmmemph{combined} influence between nuclear quantum effects and vdW interactions. The latter strongly favor the formation of a high-density liquid, whereas the inclusion of the former mitigates this by decreasing the mean hydrogen-bond (H-bond) distance. Examining the structure of water reveals that while the major fraction of molecules do in fact exhibit the traditional picture of near-tetrahedral coordination, the liquid considerably softer than previously simulations have suggested, including a much lower proportion of molecules double-donating H-bonds as well as a much larger distribution of their angles. 
\end{abstract}


\keywords{liquid water; hydrogen bonding; van der Waals interactions; nuclear quantum effects; first-principles simulations; path-integral molecular dynamics}

\maketitle




Water plays a central role in many scientific fields \cite{water_Franks}; consider, for example, its involvement in nearly all chemical, biological, and geophysical processes. Despite such broad importance, water's most basic property, its local structure at ambient conditions, characterized by the geometry of its underlying hydrogen-bond ($\text{H}$-bond) network, has remained a matter of debate for over a century \cite{XAS_water-structure_overview_Ball-Nature-2008, water-structure_review_THG-MolPhys-2010, water-struct_perspective_Nilsson-ChemPhys-2011}. The traditional view asserts that H bonds form a macroscopically-connected random network, where molecules on average exhibit near-tetrahedral coordination ($\jmmapprox 3.5$ H bonds / molecule) \cite{continuum-model_Fowler-JCP-1933} with thermal motion causing continuous topological reformations via straining and breaking \cite{H2O-struct_thermal-fluct_Stillinger-Science-1980}.
Recent small-angle X-ray scattering (SAXS) data \cite{H2O-struct_HDL-LDL-model_Huang-PNAS-2009}, however, suggests the possibility that structural polyamorphism between interconverting high-density liquid (HDL) and low-density liquid (LDL) polyamorphs may exist not only in the supercooled region \cite{HDA-LDA_HDL-LDL_review_Stanley-Nature-1998}, but at ambient conditions as well. 

Atomistic simulations have the potential to resolve these issues. Unfortunately, many challenges exist to simulating ambient water, even at an \textit{ab initio} level through density-functional theory (DFT) calculations. In fact, providing an accurate theoretical description has remained a central topic and open challenge in physical chemistry for many decades. 
Challenges arise because water is only $\jmmapprox 25$ K from the melting temperature of ice, where a variety of subtle and complex effects become important. While the structure is dominated by $\text{H}$ bonds between neighboring molecules, both van der Waals (vdW) interactions (which, in this context, refers to dispersion forces resulting from dynamical nonlocal electron correlations) and nuclear quantum effects (NQEs) influence the topology of the $\text{H}$-bond network. In fact, as discussed below, recent simulations have demonstrated that each of these effects has a \jmmemph{profound} influence on the underlying H-bond network \cite{water_DFT-D_DCACPs_Rothlisberger-JPCB-2009, water_DFT-D_NpT_Hutter-JPCB-2009, H2O_vdW-DF1_Fernandez-Serra-JCP-2011, H2O_vdW-DF2_Nilsson-JPCB-2011, H2O_NQE_Martyna-ChemPhysLett-2005, H2O_NQE_Car-PRL-2008}. 

vdW interactions are particularly important in water, due to the high polarizability of oxygen ($\text{O}$). While much weaker than $\text{H}$ bonds, they can bend those in the first coordination-shell via outer-neighbor interactions. Including vdW interactions in DFT simulations though has traditionally been problematic, as semilocal exchange--correlation density functionals (DFs) (obviously) lack the ability to describe them. As highlighted by recent simulations including empirical vdW corrections \cite{water_DFT-D_DCACPs_Rothlisberger-JPCB-2009, water_DFT-D_NpT_Hutter-JPCB-2009}, their neglect may be one of the primary reasons for such simulations to typically result in an over-structured (or supercooled) liquid with a slow self-diffusion and an overestimation of the number of $\text{H}$ bonds per molecule \cite{grossman:300, schwegler:5400}. 
Recently, however, DFs capable of calculating nonlocal correlations self-consistently \cite{PhysRevB.76.125112} have been proposed; standard ones so-called vdW-DF \cite{vdW-DF_Lundqvist-PRL-2004} and vdW-DF2 \cite{vdW-DF2_Langreth-PRBR-2010}. The form of each is similar, combining a reasonable approximation to semilocal exchange with a plasmon-pole model for correlation. Unfortunately though, while such DFs significantly improve the DFT description of small water clusters and hexagonal ice (ice $\text{I}_\text{h}$) \cite{PhysRevB.84.045116}, their application to liquid water \cite{H2O_vdW-DF1_Fernandez-Serra-JCP-2011, H2O_vdW-DF2_Nilsson-JPCB-2011} has led to the somewhat puzzling result of the formation of a HDL.

The problem with overstructuring has also been partly attributed to the lack of including NQEs \cite{H2O_NQE_Martyna-ChemPhysLett-2005, H2O_NQE_Car-PRL-2008}. These have been shown to have a large impact on the $\text{H}$-bond network, due to the fact that water is primarily in its ground vibrational-state at ambient conditions and the light $\text{H}$ mass causes zero-point motion (ZPM) to be large. These studies, however, still resulted in discrepancies with experiment, presumably because they too did not include other relevant effects, such as, perhaps, vdW interactions.

It is the purpose of this Letter to present results from simulations incorporating both vdW interactions and NQEs. It is important to note that it is not \textit{a priori} obvious how these effects should together behave, as there can be a complex interplay between intramolecular and intermolecular ZPM \cite{H2O_competing-quant-effects_Manolopoulos-JCP-2009}. The simulations were performed using  path-integral molecular dynamics (PIMD) simulations based on DFT employing vdW-DF \cite{vdW-DF_Lundqvist-PRL-2004} and vdW-DF2 \cite{vdW-DF2_Langreth-PRBR-2010} DFs; see the Methods section. These results are presented in order to provide insight into the underlying structure of water.





\begin{figure}
  \includegraphics[scale=0.5]{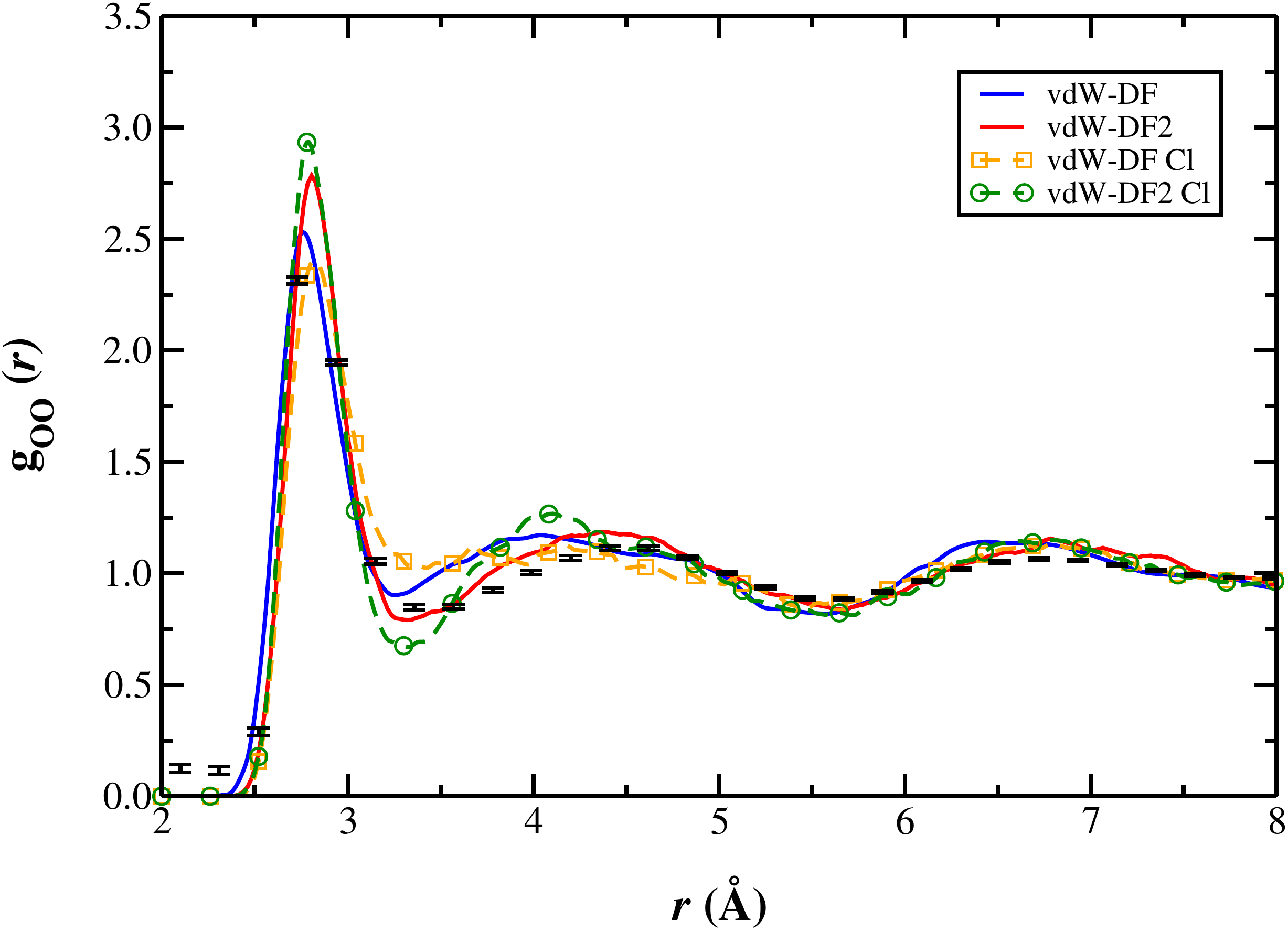}  
  \caption{Calculated $\text{O}$--$\text{O}$ PCFs, $g_\text{OO}(r)$. Error bars from PCFs derived from experimental x-ray diffraction data \cite{OO_PCF_Nilsson_JCP-2013} are also shown. The error bars for simulation results are very small, on the order of line-thickness or symbol-size, and thus are not shown. Results for classical (nuclei) simulations are denoted using ``Cl''. These latter two points are also relevant for the remaining figures, and will not be repeated.}
  \label{fig:gOO}
\end{figure}

Standard benchmarks by which the accuracy of ambient-water simulations can be judged, but, moreover, provide important insight into its angular-averaged structure, are the pair-correlation functions (PCFs); in particular, that providing insight into $\text{O}$--$\text{O}$ correlations, $g_\text{OO}(r)$; Fig.\ \ref{fig:gOO}.
%
Consistent with recent calculations \cite{H2O_vdW-DF1_Fernandez-Serra-JCP-2011, H2O_vdW-DF2_Nilsson-JPCB-2011, H2O_optB88-vdW_Galli-JCTC-2011}, the classical (nuclei) simulations demonstrate that vdW interactions cause $\text{O}$--$\text{O}$ correlations near $4$ -- $5$ \AA{} to shift inward (leading to a complete collapse of the second coordination-shell in the case of vdW-DF), relative to results derived from experimental x-ray diffraction measurements \cite{OO_PCF_Nilsson_JCP-2013}, compensated for by a slight outward shift of those in the first coordination-shell. These directions are marked in Fig.\ \ref{fig:gOO}, and taken together are indicative of the formation of (or tendency toward) a HDL \cite{PhysRevLett.84.2881}. The remarkable agreement between the vdW-DF2 and experimentally-derived results should also be noted.

Surprisingly, the inclusion of NQEs seems to mitigate these shifts, directions also marked in Fig.\ \ref{fig:gOO}. As this behavior is not seen with semilocal functionals \cite{H2O_NQE_Martyna-ChemPhysLett-2005, H2O_NQE_Car-PRL-2008}, nor with NQEs and \jmmemph{not} vdW interactions, this result strongly suggests that not only are both vdW interactions and NQEs important in describing ambient water, but there may be a heretofore unappreciated combined influence between them. This is further supported by the fact that this effect is apparent irrespective of the choice of DF, as long as election correlations are treated nonlocally \jmmemph{and} self-consistently -- see also the Supplementary Information (SI). In passing, we note that while it may appear at first that these results imply a ``hardening'' of water structure with NQEs, below it is demonstrated that this is in fact not the case.

\begin{figure}
  \includegraphics[scale=0.5]{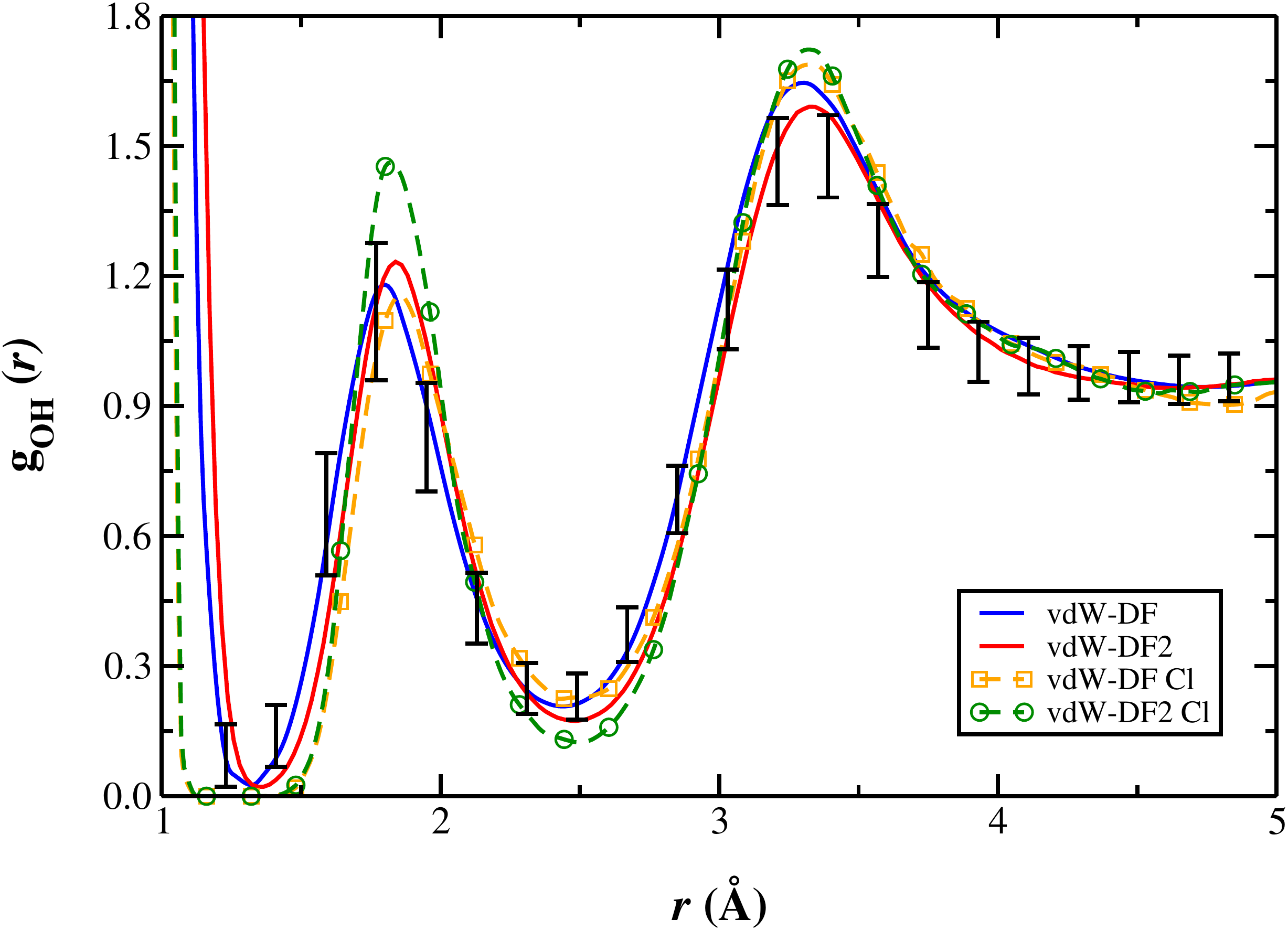}  
  \caption{Calculated $\text{O}$--$\text{H}$ PCFs, $g_\text{OH}(r)$. Error bars from PCFs derived from experimental neutron scattering data \cite{H2O-liq_neutron-data_Soper-ChemPhys-2000} are also shown.}
  \label{fig:gOH}
\end{figure}

More insight into the possibility of a combined influence between vdW interactions and NQEs is obtained via the $\text{O}$--$\text{H}$ PCF, $g_\text{OH}(r)$; Fig.\ \ref{fig:gOH}. 
Correlations near $1.3$ -- $2.4$ \AA{}, which are directly indicative of $\text{H}$ bonding, demonstrate that when vdW interactions are included, $\text{H}$ bonds increase (weaken), consistent with prior work \cite{water_DFT-D_DCACPs_Rothlisberger-JPCB-2009, water_DFT-D_NpT_Hutter-JPCB-2009, H2O_vdW-DF1_Fernandez-Serra-JCP-2011, H2O_vdW-DF2_Nilsson-JPCB-2011}. When NQEs are included, however, these correlations shift in the opposite direction (as marked in Fig.\ \ref{fig:gOH}), suggesting that the mean $\text{H}$-bond distance is decreased, or the mean strength of intermolecular interactions is increased. While this strengthening is rather small, it is apparently nonetheless enough to overcome the much weaker outer-neighbor vdW interactions \footnote{Compare also the vdW-DF and vdW-DF2 results in Figs.\ \ref{fig:gOO} and \ref{fig:gOH}, where the slight relative outward shift of $\text{O}$--$\text{H}$ correlations when using the former is likely the reason for the relatively soft second coordination-shell of $g_\text{OO}(r)$.}. It is prudent to note that Fig.\ \ref{fig:gOH} also shows there is also a greater distribution of $\text{H}$-bond distances suggesting a ``softening'' of the $\text{H}$-bond network, consistent with intuitive expectation and also previous simulations employing semilocal DFs \cite{H2O_NQE_Martyna-ChemPhysLett-2005, H2O_NQE_Car-PRL-2008}. While the effects of strengthening intermolecular interactions, yet a softer $\text{H}$-bond network may seem at odds, they are consistent with the recent findings of competing quantum effects in water \cite{H2O_competing-quant-effects_Manolopoulos-JCP-2009}, mentioned above, where intermolecular ZPM and tunneling effects soften the $\text{H}$-bond network, while intramolecular ZPM results in strengthened intermolecular interactions, which arise from an increased dipole moment. Again note the level of agreement between the vdW-DF2 results and those experimentally-derived.

\begin{figure}
  \includegraphics[scale=0.5]{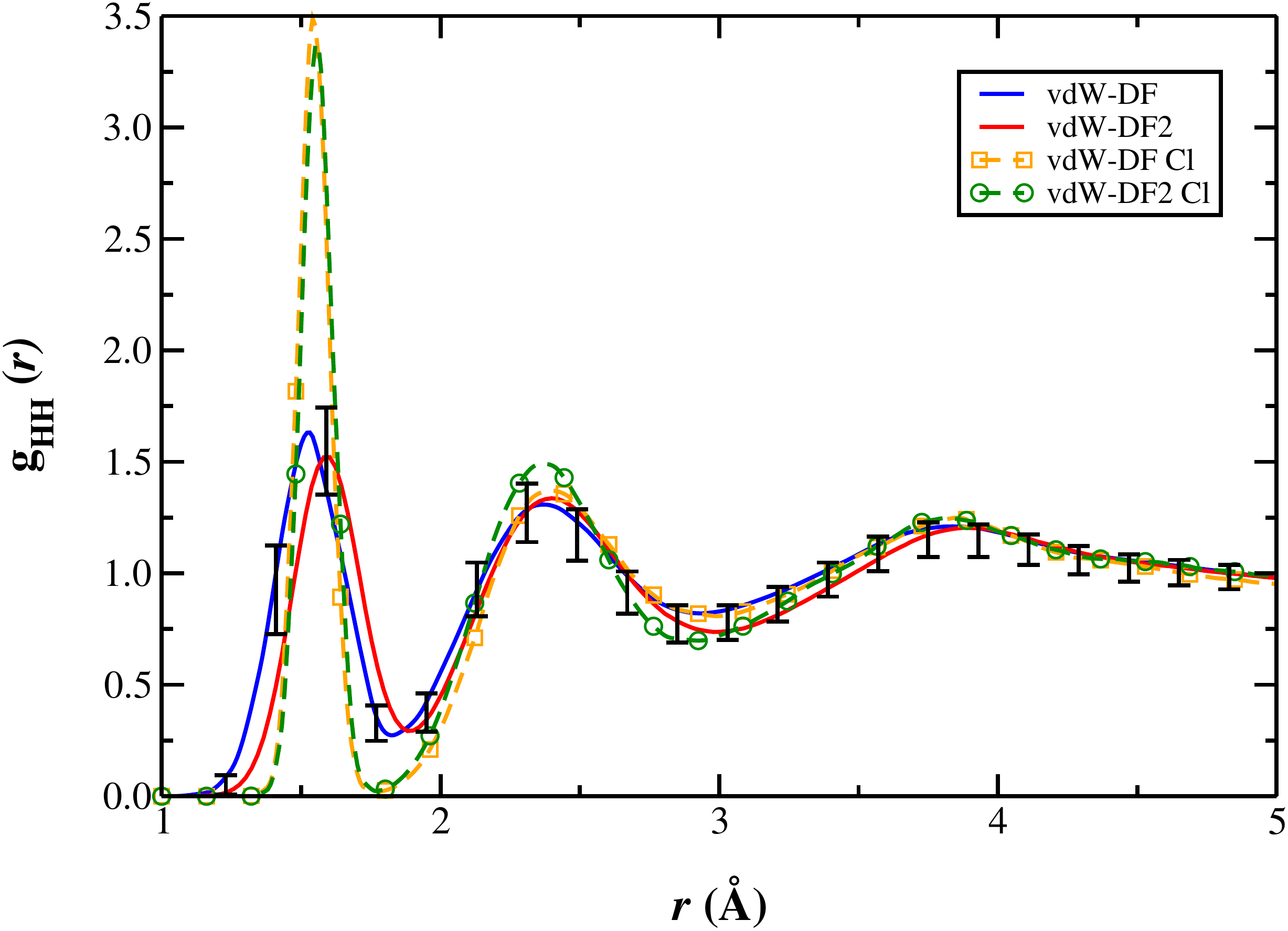}  
  \caption{Calculated $\text{H}$--$\text{H}$ PCFs, $g_\text{HH}(r)$. Error bars from PCFs derived from experimental neutron scattering data \cite{H2O-liq_neutron-data_Soper-ChemPhys-2000} are also shown.}
  \label{fig:gHH}
\end{figure}

While $g_\text{OO}(r)$ and $g_\text{OH}(r)$ provide the primary insight into the combined influence between NQEs and vdW interactions, for a complete description of the angular-averaged structure, we show the $\text{H}$--$\text{H}$ PCF, $g_\text{HH}(r)$, in Fig.\ \ref{fig:gHH}.
From this, insight into both the intramolecular potential energy surface of an isolated molecule in the condensed-phase environment and the relative orientations of neighboring molecules (given by the first and second peaks, respectively), can be inferred. For example, it can be seen that the classical simulations poorly describe intramolecular motion, which is significantly improved with NQEs, as expected, because water is primarily in its ground vibrational-state at ambient conditions and NQEs are large, also mentioned above.



\begin{figure}
  \includegraphics[scale=0.3]{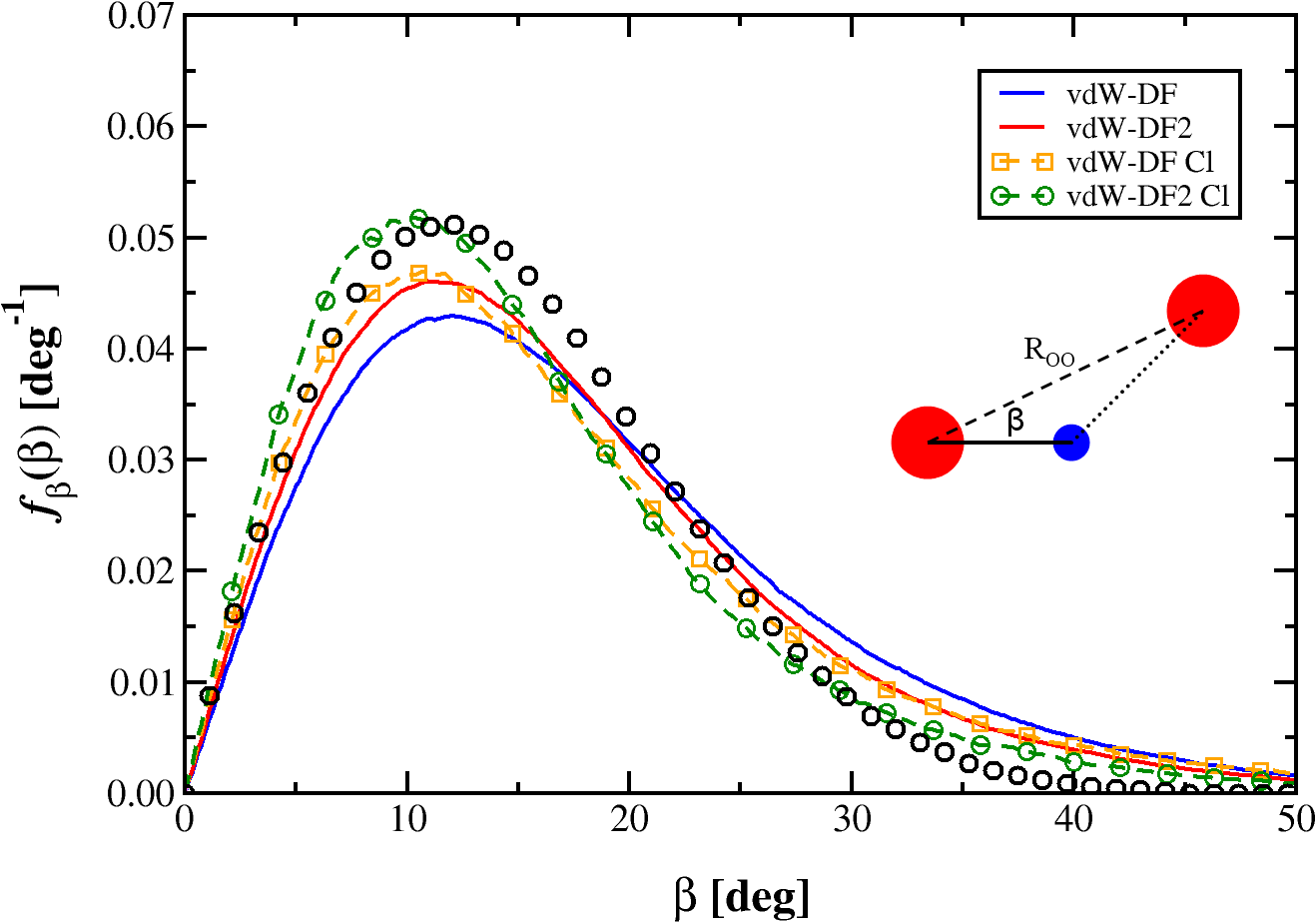}
  \caption{Calculated probability distribution functions $\fbeta$ of the $\text{H}$-bond angle $\beta$ (shown in the inset), compared to experimentally-derived results \cite{H-bond_geom_exp_Halle-PRL-2003}, shown as open black circles.}
  \label{fig:H-bond_prob_dist}
\end{figure}

While the PCFs give important insight into the angular-averaged structure of water, they do not provide information about angular correlations that are necessary to understand local geometries. One measure that does give such information though is the probability distribution function $\fbeta$ of $\text{H}$-bond angles, which quantifies the probability of finding a $\text{H}$ bond at an angle $\beta$, the latter considered to be that between the two $\text{O}$ atoms and the covalent $\text{O}$--$\text{H}$ direction (see the inset of Fig.\ \ref{fig:H-bond_prob_dist}).
Note that the definition of a $\text{H}$ bond is not rigorous, with many possibilities existing \cite{H-bond_def_Skinner-JCP-2007}, and so in order to make a comparison to existing results derived by experiment \cite{H-bond_geom_exp_Halle-PRL-2003}, in this case, we consider each $\text{H}$ atom to participate in a $\text{H}$ bond, the latter found by simply determining the two nearest $\text{O}$ atoms (a more sophisticated definition is utilized below). Figure \ref{fig:H-bond_prob_dist} shows that, on average, $\text{H}$ atoms are shifted slightly outward from the linear $\text{O}$--$\text{O}$ direction, positioned essentially in a ``cone'' of angles around the donating $\text{O}$ atom (peaked near $11.6$\textdegree). The inclusion of NQEs shifts the peak angle even farther outward, as well as increases the angular distribution, both demonstrative of a softening of the $\text{H}$-bond network, confirming some of the remarks made above.


In addition to angular correlations, it is also important to consider actual bonding configurations. A quantitative measure can be made by looking at the relative proportions of $\text{H}$ bonds donated by each $\text{O}$ atom, the latter each of which can be labeled as non-donating (ND), single-donating (SD), or double-donating (DD) (higher proportions, resulting from $\text{H}_3\text{O}^{+}$, for example, are negligibly small); Table \ref{Tb:no_H-bonds_microscopic}.
\begin{table}
  \caption{Calculated percentages of ND, SD, and DD $\text{O}$ atoms (in reference to $\text{H}$ bonds), in comparison to theoretical liquid-state spectroscopy results \cite{H-bond_dist_Skinner-PNAS-2007}, as described in the text.}
  \label{Tb:no_H-bonds_microscopic}
  \begin{ruledtabular}
    \begin{tabular}{c c c c}
                                     & ND         & SD         & DD         \\
      \hline
      \hline
      Theor.\ Liq.-state Spect.\                            & $3$        & $27$       & $70$      \\
      \hline
      vdW-DF                         & $6$        & $39$       & $55$       \\
      \hline
      vdW-DF2                        & $3$        & $32$       & $65$       \\
      \hline
      vdW-DF Cl                      & $5$        & $42$       & $53$      \\
      \hline
      vdW-DF2 Cl                     & $2$        & $29$       & $69$      \\
    \end{tabular}
  \end{ruledtabular}
\end{table}
In what follows, a $\text{H}$-bond using the geometric criterion of Ref.\ \onlinecite{H2O-struct_SD-H-bond-model_Wernet-Science-2004}: one is considered to exist between two $\text{O}$ atoms if the separation of the latter ($\text{R}_\text{OO}$) satisfies $\text{R}_\text{OO} \leq -0.00044 \beta^2 + 3.3$ \AA{}, with $\beta$ in degrees. Note that this quantity is shown in the inset of Fig.\ \ref{fig:H-bond_prob_dist}, and in this case, precisely corresponds to the aforementioned cone of angles around the donating $\text{O}$ atom for which a $\text{H}$ atom may be found. Note also that Table \ref{Tb:no_H-bonds_microscopic} shows theoretical liquid-state spectroscopy results \cite{H-bond_dist_Skinner-PNAS-2007} as a reference, which are based on  calculated IR and isotropic Raman spectra fit to experiment (a procedure \jmmemph{assumed} to be independent of the underlying computational method), as definitive experimental conclusions (interpretations) are still matters of debate \cite{XAS_water-structure_overview_Ball-Nature-2008, water-structure_review_THG-MolPhys-2010, water-struct_perspective_Nilsson-ChemPhys-2011}. 

Table \ref{Tb:no_H-bonds_microscopic} provides important insight into the actual structure of ambient water. It suggests that just under two-thirds of molecules actively exhibit the traditional picture of near-tetrahedral (DD) coordination (see the vdW-DF2 results). This is considerably softer than current thinking \cite{water-structure_review_THG-MolPhys-2010} (but which in large part has been based on previous \textit{ab initio} simulations -- e.g., Refs.\ \onlinecite{H2O_NQE_Martyna-ChemPhysLett-2005} and \onlinecite{H2O_NQE_Car-PRL-2008}) that suggest such coordination is closer to $\jmmapprox 80\%$. Nonetheless, these results agree quite well with the theoretical liquid-state spectroscopy results \cite{H-bond_dist_Skinner-PNAS-2007}, and in fact resolve discrepancies between the aforementioned previous \textit{ab initio} simulations \cite{H2O_NQE_Martyna-ChemPhysLett-2005} and sophisticated empirical models \cite{TTM2p1-F-model_quantum-effects_Voth-JCP-2007}. 

While the percentage of DD species is relatively low, they do nonetheless continue to form the majority, indicative of average near-tetrahedral coordination. What is at first surprising is that the inclusion of NQEs is found to have only a small impact on the results, as a reduction in DD species is expected (e.g., empirical simulations \cite{TTM2p1-F-model_quantum-effects_Voth-JCP-2007} suggest a $\jmmapprox 15\%$ reduction at ambient temperature). However, this is further consistent with a scenario where there is a combined influence between NQEs and vdW interactions; while the former alone likely reduce the proportion of DD species, in the presence of vdW interactions, they mitigate the large reduction already caused, and in the end, seemingly having little overall impact.



In conclusion, the inclusion of both NQEs and vdW interactions into \textit{ab initio} DFT simulations provides a remarkably accurate description of ambient water (the vdW-DF2 results, in particular). Excellent agreement with experiment (or results derived therefrom) was obtained in all cases considered.
The level of agreement should be especially compared to the plethora of results from previous simulations, which have resulted in severe over-structuring or the formation of a HDL without resort to \textit{ad hoc} or empirical corrections; see Refs.\ \onlinecite{grossman:300, schwegler:5400,H2O_vdW-DF1_Fernandez-Serra-JCP-2011, H2O_vdW-DF2_Nilsson-JPCB-2011}, for example.
Furthermore, the results suggest that a combined influence between NQEs and vdW interactions impacts the structure of ambient water; vdW interactions favor the formation of a HDL, while NQEs mitigates this by strengthening the mean intermolecular interaction. The importance and influence of such subtle effects (combined NQEs and vdW interactions) on the structure (and likely behavior) of ambient water suggests some clear caveats to future simulations, in particular \textit{ad hoc} or empirical corrections used to account for them (see the SI). Looking forward, it is quite possible that these subtle effects, and the competition that they exhibit, are not only important in water, but also other liquids containing $\text{H}$-bonds.


\

\noindent
\textbf{Methods}

Simulations were performed using the Quantum ESPRESSO \textit{ab initio} DFT code \cite{QE-2009}. Distribution versions of ultrasoft pseudopotentials generated with the Perdew-Burke-Ernzerhof (PBE)  \cite{PBE_exch-correl_PRL-1996} DF were used for both $\text{O}$ and $\text{H}$. A plane-wave basis set with kinetic-energy and charge-density cutoffs of $30$ and $240$ Ry, respectively, was found to adequately converge forces for the molecular dynamics. Periodic boundary conditions were enforced automatically by the basis set and $\Gamma$-point Brillouin-zone sampling. A number of simulations were performed with various DFs (semilocal, in some cases with an empirical vdW correction, as well as those capable of treating nonlocal correlations self-consistently), as discussed in the main text or SI. For classical simulations, statistics were collected in the canonical ensemble using an Andersen thermostat \cite{Andersen_thermostat_JCP-1980} to control the temperature.
Quantum statistics were generated using an accelerated PIMD approach based on a generalized Langevin equation \cite{PI+GLE_Parrinello-JCP-2011}, with $8$ imaginary-time slices (which was found to be adequate to converge both the quantum kinetic and potential energies). For both types of simulation, a time step of $10$ a.u.\ was used to propagate the trajectories.

To represent ambient water, $64$ $\text{H}_2\text{O}$ molecules (with atomic masses of $1.00794$ and $15.9994$ for $\text{H}$ and $\text{O}$, respectively) were simulated at a density of $1$ $\text{g} / \text{cm}^3$ and temperature of $300$ K. An estimate of the equilibrium density for vdW-DF2 to justify this choice (also used for other functionals) is provided in the SI. Equilibration of the liquid was performed by first (semi-)melting ice $\text{I}_\text{h}$ at $400$ K and performing classical simulations using PBE for $\jmmapprox 8$ ps. Further equilibration was then performed for a further few ps at the PIMD-level, followed by many ps of statistics collection. For vdW-DF2, for example, precise times were $\jmmapprox 7$ ps equilibration and a further $8$ ps of statistics collection, for a total simulation time of $23$ ps.


\begin{acknowledgments}
We thank David M.\ Ceperley and Eric Schwegler for insightful discussions and providing helpful comments on an early draft of this manuscript. J.\ M.\ M.\ was supported by DOE Grant DE-NA$0001789$. M.\ A.\ M.\ was supported by the DOE at Lawrence Livermore National Laboratory under Contract DE-AC52-07NA27344. This research was also supported in part by the NSF through XSEDE resources provided by NICS under grant number TG-MCA93S030.
\end{acknowledgments}


\

\noindent
\textbf{Associated Content}

\

\noindent
\textbf{Supporting Information.} Calculation of the equilibrium density of vdW-DF2; results for semi-local DFs, with and without empirical vdW corrections, as well as additional vdW-DFs; caveats to future simulations. This material is available free of charge via the Internet at http://pubs.acs.org.


\end{document}